\documentclass[aps,pre,reprint,groupedaddress,superscriptaddress]{revtex4-1}

\usepackage[pdftex]{graphicx}
\usepackage{amssymb,amsfonts,amsmath}
\usepackage{grffile}

\begin{document}



\title{Two-cluster solutions in an ensemble of generic limit-cycle 
oscillators with periodic self-forcing via the mean-field}


\author{Lennart Schmidt}
\affiliation{Physik-Department, Nonequilibrium Chemical Physics, Technische Universit\"{a}t M\"{u}nchen,
  James-Franck-Str. 1, D-85748 Garching, Germany}
\affiliation{Institute for Advanced Study - Technische Universit\"{a}t M\"{u}nchen,
  Lichtenbergstr. 2a, D-85748 Garching, Germany}

\author{Katharina Krischer}
\email[]{krischer@ph.tum.de}
\affiliation{Physik-Department, Nonequilibrium Chemical Physics, Technische Universit\"{a}t M\"{u}nchen,
  James-Franck-Str. 1, D-85748 Garching, Germany}


\date{\today}

\begin{abstract}

We study two-cluster solutions of an ensemble of generic limit-cycle 
oscillators in the vicinity of a Hopf bifurcation, i.e. 
Stuart-Landau oscillators, with a nonlinear global coupling. This 
coupling leads to conserved mean-field 
oscillations acting back on the individual oscillators as a forcing. A 
reduction to two effective equations makes a linear stability 
analysis of the cluster solutions possible. These equations exhibit a $\pi$-rotational 
symmetry, leading to a complex bifurcation structure and a wide 
variety of solutions. In fact, the principal bifurcation structure
resembles that of a 2:1 resonance tongue, while inside the tongue we
observe an 1:1 entrainment.

\end{abstract}


\maketitle


\section{Introduction}
Cluster formation is a well known phenomenon in systems of coupled
oscillators. It arises in discrete systems of individual units
\cite{Kaneko_PhysicaD_1990, Okuda_PhysicaD_1993, Nakagawa_PhysicaD_1994} and in
spatially extended oscillatory media \cite{Vanag_Nature_2000,
  Vanag_JPCA_2000, Mikhailov_PhysicsReports_2006, Lin_PRE_2004,
  Kaira_PRE_2008}. The common property of clusters in these systems is
that the oscillators separate into distinct groups having the same 
properties within. The oscillations in the
different groups are then phase shifted with respect to each other. In
the symmetrical phase cluster state, the phase shifts for $n$ clusters
are given by $2 \pi m/n$ \cite{Okuda_PhysicaD_1993, Lin_PRE_2004,
  Kaira_PRE_2008}, where $m=1,2, \dots, n-1$. Here, we will treat 
  two-cluster solutions, exhibiting more complex than simple periodic
  dynamics. In many cases the
amplitude variations in these states are very small and the dynamics
can be approximated by phase models. However, as one prominent counter
example, we present a type of
clusters, so-called type II clusters \cite{Varela_PCCP_2005}, where essential variations in the amplitudes occur. They
have been described in Refs.~\cite{Varela_PCCP_2005,
  Miethe_PRL_2009, GarciaMorales_PRE_2010, Bertram_JPCB_2003}.
In this state the clusters are a modulation of a homogeneous
oscillation, as visible in Fig.~\ref{fig:clusters}a.

The photoelectrodissolution of n-type silicon \cite{Miethe_PRL_2009,
  Schoenleber_NJP_2014} is an experimental system exhibiting this type
of clustering. Many of the spatio-temporal dynamics of this system can
be modelled with a complex Ginzburg-Landau equation (CGLE) with a nonlinear
global coupling \cite{GarciaMorales_PRE_2010, Schmidt_Chaos_2014}. As
the essential ingredient for the dynamics is this nonlinear global
coupling \cite{Schmidt_arXiv_2014}, we drop the diffusive coupling of the
CGLE, rendering a mathematical treatment
of the cluster solutions possible. Thus, in this Article we are
dealing with an ensemble of Stuart-Landau oscillators, coupled via a
nonlinear global coupling. As we will see, this coupling leads to a 
conserved periodic mean-field oscillation that acts back on the 
individual oscillators as a forcing. Then, we reduce the full set of equations to two
effective equations describing the case of clustering with two
groups. We show that we end up with an equation possessing the same
(symmetry) properties as the resonantly forced CGLE near a 2:1 
resonance, which also exhibits cluster formation. The symmetry of this 
equation leads to a very complex bifurcation diagram and therefore to
a wide variety of different dynamical states, in line with results on
periodically forced oscillators near a 2:1 resonance
\cite{Vance_JCP_1989, Vance_Chaos_1991}, with one exception: inside the
locking region we observe an 1:1 entrainment, despite the bifurcation
structure of a 2:1 resonance.

\section{Stuart-Landau oscillators with a conservation law}
Our model consists of $N$ Stuart-Landau oscillators, each of the form
\cite{Nakagawa_PhysicaD_1994, Kuramoto_2003, Daido_PRL_2006}

\begin{equation}
  \frac{\mathrm d}{\mathrm dt} W_k = W_k - (1+i c_2) \left| W_k
  \right|^2 W_k \ , \quad k = 1, \dots, N \ ,
\end{equation}

coupled via a nonlinear global coupling \cite{Miethe_PRL_2009,
  Schmidt_Chaos_2014}:

\begin{align}
  \frac{\mathrm d}{\mathrm dt} W_k = &W_k - (1+i c_2) \left| W_k
  \right|^2 W_k \notag \\
  &- (1+i \nu) \left< W \right> + (1+i c_2)
  \left< \left| W \right|^2 W \right> \ .
\label{eq:SL_ensemble}
\end{align}

Here $\left< \dots \right>$ describes the arithmetic mean over the
oscillator population, i.e. $\left< W \right> = \sum_{k=1}^N W_k /
N$. Taking the average of the whole equation yields for the dynamics
of the mean value

\begin{equation}
  \frac{\mathrm d}{\mathrm dt} \left< W \right> = - i \nu \left<
    W \right> \quad \Rightarrow \quad \left< W \right> = \eta
  e^{-i \nu t} \ .
\label{eq:conservation_law}
\end{equation}

Therefore, we are dealing with a globally coupled population of
Stuart-Landau oscillators with a conservation law for the mean-field
oscillation. This conservation is an important property of the
dynamics of the experimental silicon system. Here, we achieve it by
the specific design of our coupling function. This mean-field 
oscillation also acts as an intrinsic self-forcing on the 
individual oscillators. We will see that this indeed leads to a
so-called Arnold tongue, a tongue-shaped region in which the
oscillations are entrained to the driving.
In general, the dynamics of the oscillator population, Eqs.~\eqref{eq:SL_ensemble}, is determined by three
parameters, namely $c_2$, $\nu$ and $\eta$. 

Equations~\eqref{eq:SL_ensemble} are equivariant to the direct product
$\mathbf S_N \times S^1$ of the symmetry group $\mathbf S_N$ of
permutations of $N$ elements and the circle group $S^1$, describing
the global phase invariance. The equivariance to $\mathbf S_N$ is
obvious, as a permutation of the indices in
Eqs.~\eqref{eq:SL_ensemble} leaves the whole set of equations
invariant. Nevertheless, particular solutions are not required to
possess the full $\mathbf S_N \times S^1$ symmetry, only all 
solutions together exhibit it \cite{Golubitsky_2003}.

\begin{figure}[ht]
  \centering
  \includegraphics[width=8.5cm]{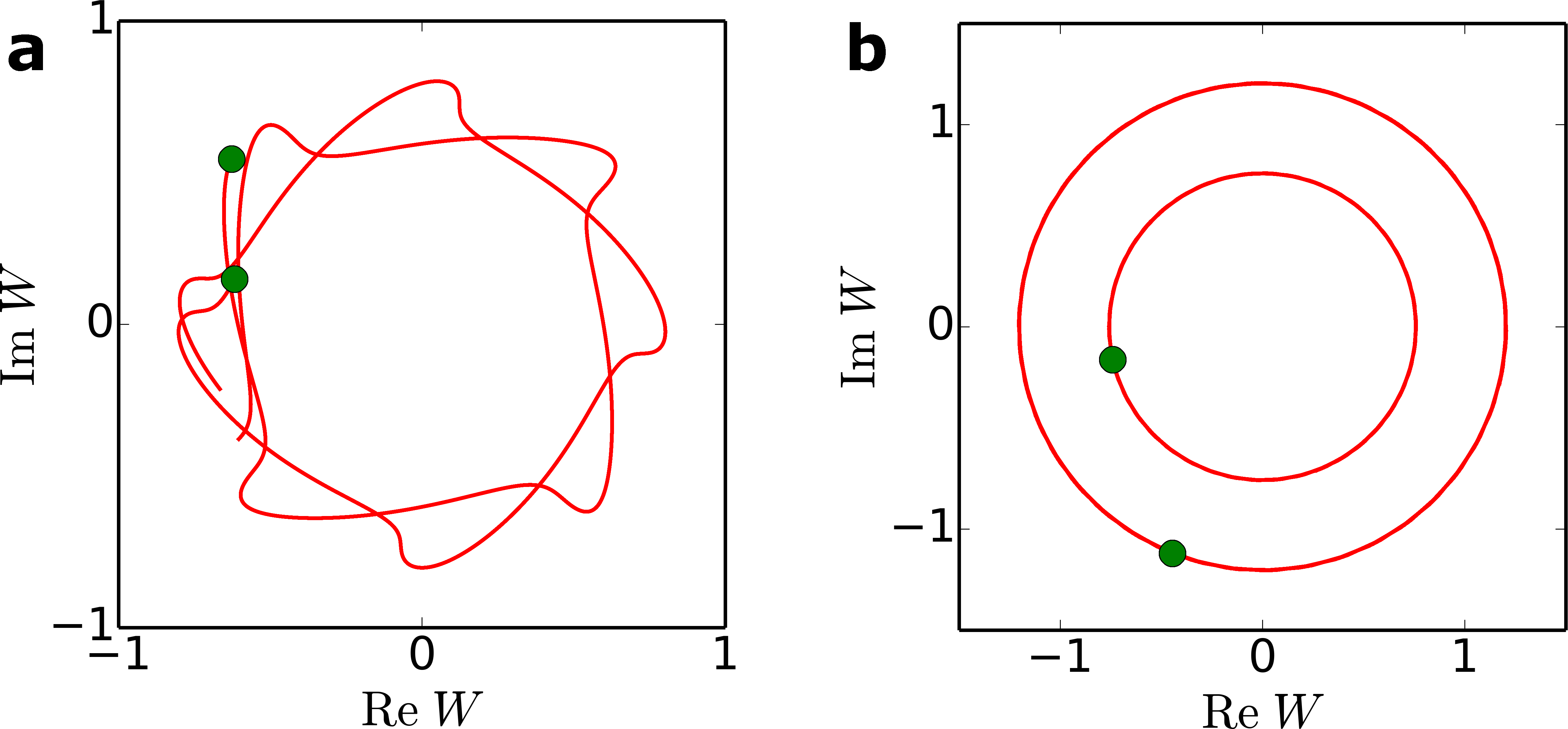}
  \caption{Cluster dynamics in the Stuart-Landau ensemble. Lines
    describe the trajectories and dots mark the positions in a
    snapshot. (a) Modulated amplitude clusters for
    $c_2 = -0.6$, $\nu = 0.1$ and $\eta = 0.7$. Here, the subgroups
    perform additional oscillations around their mean-field $\eta
    \exp(-i\nu t)$ given in Eq.~\eqref{eq:conservation_law}.
    (b) Amplitude clusters for $c_2 = -0.6$,
    $\nu = -1.5$ and $\eta = 0.9$. The main differences between the two
    groups are the different radii of their respective limit
    cycles. The phase shift is much smaller than $\pi$.}
  \label{fig:clusters}
\end{figure}

We numerically solved
Eqs.~\eqref{eq:SL_ensemble} using an implicit Adams method with
timestep $dt = 0.01$. 
For certain parameter regimes the whole
population divides into two subgroups of size $N_1$ and $N_2$ with
$N_1 + N_2 = N$. Thus, the full symmetry is reduced to $\mathbf
S_{N_1} \times \mathbf S_{N_2} \times S^1 \subseteq \mathbf S_N \times
S^1$. For $\eta > 0$ one then observes modulated amplitude and
amplitude clusters as shown in Figs.~\ref{fig:clusters}a and
b, respectively.

In the modulated amplitude cluster state the
subgroups oscillate, in addition to the mean-field oscillation, around
their mean field. This leads to a repeated passing by each other of the
subgroups in the complex plane. Similar states were observed in
continuous systems in Refs.~\cite{Vanag_JPCA_2000, Pollmann_CPL_2001,
  Bertram_JPCB_2003,Varela_PCCP_2005, Miethe_PRL_2009, GarciaMorales_PRE_2010}.
In the amplitude cluster state the two groups
oscillate on different limit cycles separated by an amplitude
difference, while the phase shift is much smaller than $\pi$
\cite{Daido_PRL_2006}. In the next section, in order to treat these
solutions mathematically, we reduce the full set
of $N$ equations in Eqs.~\eqref{eq:SL_ensemble} to two effective equations
modelling the two subgroups.

\section{Modulated amplitude clusters in the two-groups reduction}

We will now focus on the modulated amplitude clusters as presented in
Fig.~\ref{fig:clusters}a. As visible from the figure, the ensemble
splits into two groups, each performing amplitude-modulated
oscillations in the complex plane. To analyze these dynamics, we
reduce the $N$ equations of the Stuart-Landau ensemble, Eqs.~\eqref{eq:SL_ensemble}, to two
effective equations. Therefore, we assume two groups $W_1$ and $W_2$, each
synchronized, with sizes $N_1$ and $N_2$, respectively. The average
over the entire ensemble is then given by

\begin{equation}
  \left< W \right> = \frac 1N \left( N_1 W_1 + N_2 W_2 \right) \ ,
\end{equation} 

and analogously for $\left< \left| W \right|^2 W
\right>$. Inserting these expressions into Eqs.~\eqref{eq:SL_ensemble} results in 

\begin{align}
  \frac{\mathrm d}{\mathrm dt} W_1 = &\left( 1 - (1+i\nu)
    \frac {N_1}N \right) W_1 \notag \\
  &- (1+ic_2) \left( 1 - \frac{N_1}N \right)
  \left| W_1 \right|^2 W_1 \notag \\
  &-(1+i\nu) \frac{N_2}N W_2 + (1+ic_2) \frac{N_2}N \left| W_2
  \right|^2 W_2 \ ,
\label{eq:two_groups_reduction}
\end{align}

where the same holds for $W_2$ with indices 1 and 2
interchanged. Thus, we reduced the set of $N$ equations to two effective
equations and can now perform a linear stability analysis of the
synchronized state. By setting $W_1 = W_2$ we obtain

\begin{equation}
\frac{\mathrm d}{\mathrm dt} W_1 = \frac{\mathrm d}{\mathrm dt} W_2 =
-i \nu W_1 = -i \nu W_2 \ ,
\end{equation}

and thus

\begin{equation}
  W_1 = W_2 = \eta e^{-i\nu t} = W_0 \ ,
\end{equation}

as expected. Since the conservation law, Eq.~\eqref{eq:conservation_law}, still has to be fulfilled,
the synchronized solution is given by $W_0$. We define deviations $w_1$ and $w_2$ from $W_0$
via $W_1 = W_0(1+w_1)$ and $W_2 = W_0(1+w_2)$. To fulfill the
conservation law, 

\begin{equation}
  \frac 1N \left( N_1 w_1 + N_2 w_2 \right) = 0
\label{eq:cond_w}
\end{equation}

holds. For symmetric cluster states $N_1 = N_2 = N/2$ one obtains for
$w_1$ and $w_2$, when using the condition in Eq.~\eqref{eq:cond_w},

\begin{align}
  \frac{\mathrm d}{\mathrm dt} w_1 &= \left( \mu + i \beta \right) w_1
  - (1+ic_2) \eta^2 \left( \left| w_1 \right|^2 w_1 + w_1^* \right) \
  , \notag \\
  w_2 &= -w_1 \ ,
\label{eq:two_groups_reduction_w}
\end{align}

where $\mu = 1 - 2\eta^2$ and $\beta = \nu - 2 \eta^2 c_2$. 

Note here already that the equation for $w_1$ is a forced CGLE near a 
2:1 resonance \cite{CoulletEmilsson_PhysicaD_1992} without the
diffusive coupling, which is a result 
of the self-forcing in the system.
The synchronized solution $W_0$ possesses the symmetry $\mathbf S_2
\times S^1$ in the present two-oscillators description. A bifurcation
with emanating solution branches exhibiting the reduced symmetry
$S^1$ (separation into two subgroups) has to have the following
symmetry property: the sum of the two solutions $W_1 + W_2$ is
required to possess the full symmetry $\mathbf S_2 \times
S^1$ ($W_1$ is on one of the solution branches and $W_2$ on
another). Therefore, the symmetry breaking parts $w_1$ and $w_2$ 
have to cancel each other, i.e. $w_1 = - w_2$. This symmetry
condition is fulfilled by three types of bifurcations, namely the
pitchfork, the Hopf and the period doubling bifurcations. In case of
the Hopf and the period doubling bifurcations the two
solution branches are phase shifted by $\pi$.
We will see that we indeed find the pitchfork and the Hopf bifurcation in the following linear
stability analysis of the synchronized state, which is given by $w_1 = w_2 = 0$. The linear
stability of this state is determined by

\begin{align}
  \frac{\mathrm d}{\mathrm dt}
  \begin{pmatrix}
    w_1 \\
    w_1^*
  \end{pmatrix}
  =
  \begin{pmatrix}
    \mu + i \beta & -(1+ic_2)\eta^2 \\
    -(1-ic_2)\eta^2 & \mu - i \beta
  \end{pmatrix}
  \cdot
  \begin{pmatrix}
    w_1 \\
    w_1^*
  \end{pmatrix} \ .
\end{align}

The eigenvalues of the Jacobian matrix are given by

\begin{equation}
  \lambda_\pm = 1-2\eta^2 \pm \sqrt{\eta^4 \left( 1-3c_2^2 \right) + 4 \nu
      c_2 \eta^2 - \nu^2} \ .
\label{eq:eigenvalues}
\end{equation}

\begin{figure}[t]
  \centering
  \includegraphics[width=8.5cm]{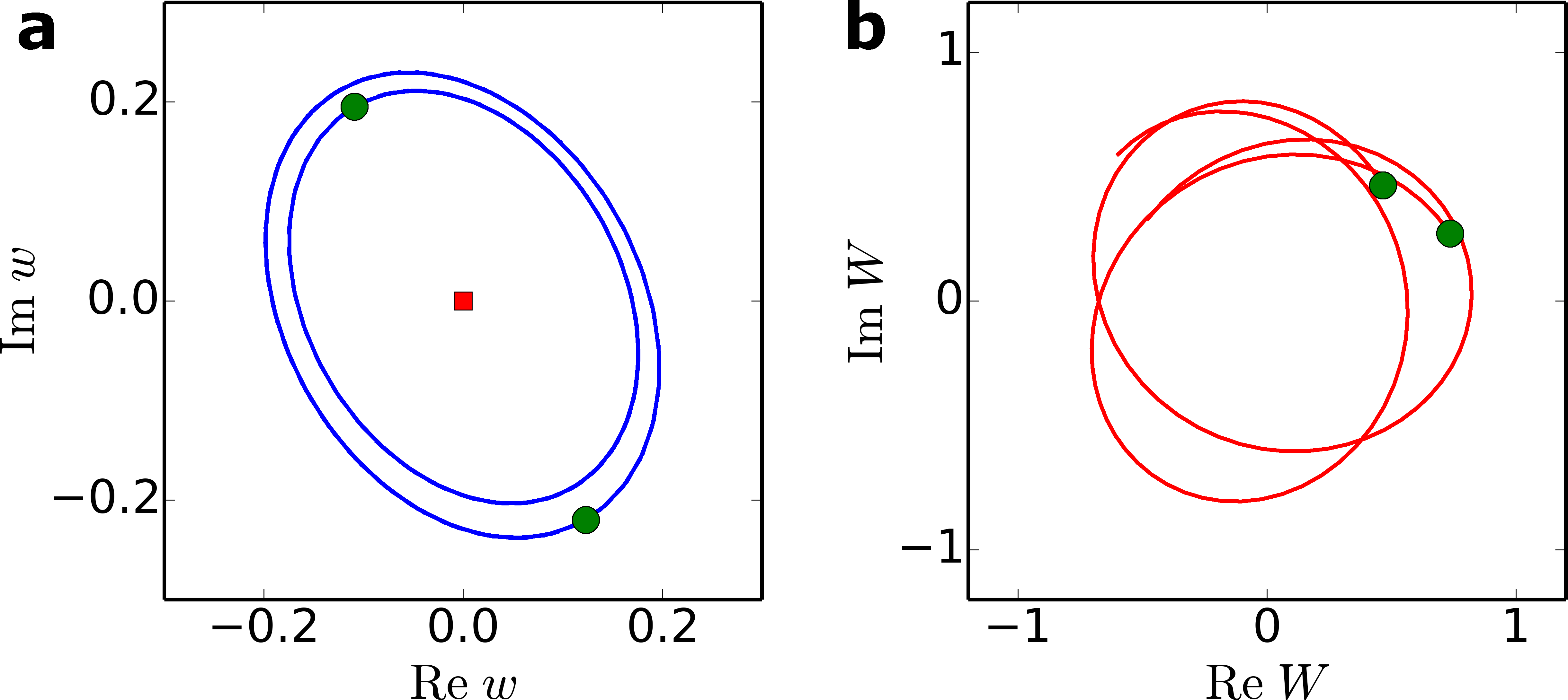}
  \caption{Emergence of modulated amplitude clusters in the full
    ensemble for $c_2 = -0.6$, $\nu = 1.2$ and $\eta = 0.7$. (a) Two limit
    cycles in anti phase and the fixed point at $\eta$ (red square) in
    the rotating frame of $w_k$ defined via $W_k = W_0(1+w_k)$. The
    limit cycles have different radii as the whole population is divided
    into subgroups with $N_1 \neq N_2$. (b) Dynamics in the original system.}
\label{fig:two_groups_cluster_emergence}
\end{figure}

Thus, we find a secondary Hopf bifurcation in
this system at $\eta = \eta_H = 1/\sqrt 2$ for $\left(1 - 3c_2^2 \right)/4 + 2 \nu c_2 - \nu^2 < 0$. This Hopf bifurcation is
the origin of the modulated amplitude clusters shown in Fig.~\ref{fig:clusters}a. 
In order to visualize this, we use the ansatz $W_k = W_0(1+w_k)$ in
the full system (Eqs.~\eqref{eq:SL_ensemble}) for the analysis of
simulation results. An example for the dynamics of $w_k$ in case of
the modulated amplitude cluster state is shown in
Fig.~\ref{fig:two_groups_cluster_emergence}a. 

One can clearly identify
the two limit cycles of the two subgroups (blue solid lines). The
green dots mark a snapshot of the dynamics. In this
reference frame the phase shift between the two groups is given by
$\pi$. The two limit cycles
are not identical, since the full system is divided into two groups with
different sizes, i.e. $N_1 \neq N_2$. This results in different radii
of the limit cycles in order to fulfill the condition in
Eq.~\eqref{eq:cond_w} and thus to fulfill the conservation law in
Eq.~\eqref{eq:conservation_law}. The red square marks the position of
the synchronized solution. These observations confirm the result of
the two-groups analysis that the modulated amplitude clusters arise in
a secondary Hopf bifurcation. Transforming the system back to $W_k$,
one obtains the dynamics shown in Fig.~\ref{fig:two_groups_cluster_emergence}b.

Using the eigenvalues in Eq.~\eqref{eq:eigenvalues} we can determine
the Hopf frequency $\omega_H$ to be

\begin{equation}
  \omega_H = \mathrm{Im} \left(\sqrt{\eta^4\left(1-3 c_2^2 \right)
      + 4 c_2 \eta^2 \nu -\nu^2} \right) \ .
\label{eq:hopf_freq}
\end{equation}

\begin{figure}[t]
  \centering
  \includegraphics[width=8.5cm]{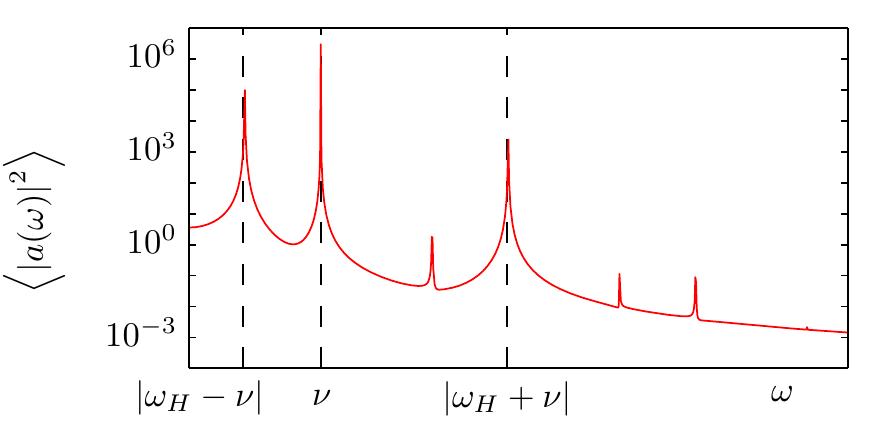}
  \caption{Cumulative power spectrum for the full system at parameter
    values  $c_2 = -0.6$, $\nu = 1.2$, $\eta = 0.7$. The major peaks in this spectrum can be traced back to
    linear combinations of the Hopf frequency $\omega_H$ in
    Eq.~\eqref{eq:hopf_freq} and the frequency of the mean-field
    oscillation $\nu$ as indicated by the vertical lines (see text and Eq.~\eqref{eq:two_groups_resonances}).}
\label{fig:two_groups_Scum}
\end{figure}

Next, we investigate the frequencies occurring in the dynamics
in the original frame. Therefore, we calculate the cumulative power
spectrum. To obtain this, one first has to Fourier transform all
individual time series $\mathrm{Re} \ W_k$ of the oscillators and then average the
resulting squared amplitudes $\left| a_k(\omega) \right|^2$, where $k$
is the oscillator index. It is thus given by $S(\omega) = \left<
  \left| a(\omega) \right|^2 \right>$.
An exemplary cumulative power spectrum for the dynamics in the
modulated amplitude cluster state (in the full system) is shown in
Fig.~\ref{fig:two_groups_Scum} and it exhibits several peaks.

The strongest peak is at the frequency $\nu$ of the
average-oscillation. As we will show in what follows, the next two
highest peaks are given by $\pm (\nu - \omega_H)$ and $\pm (\nu +
\omega_H)$ as indicated by vertical lines in the figure. 

In the vicinity of the Hopf bifurcation, the limit-cycle solution for
$w_1$ in Eq.~\eqref{eq:two_groups_reduction_w} is given by

\begin{equation}
  w_1 = w^0_+ e^{i\omega_H t} + w^0_- e^{-i\omega_H t} \ ,
\end{equation}

where $w^0_\pm$ are complex-valued constants. In the original frame
this results in 

\begin{align}
  W_1 &= \eta e^{-i\nu t} \left( 1 + w^0_+ e^{i\omega_H t} + w^0_-
    e^{-i\omega_H t} \right) \ , \notag \\
  W_2 &= \eta e^{-i\nu t} \left( 1 - w^0_+ e^{i\omega_H t} - w^0_-
    e^{-i\omega_H t} \right) \ .
\end{align}

Thus, we obtain frequency contributions in the cumulative power
spectrum at

\begin{align}
  &\pm \nu \quad \left( \propto \eta^2 \right) \ , \notag \\
  &\pm (\nu - \omega_H) \quad \left( \propto \left( \eta \omega^0_+ \right)^2 \right) \
  , \notag \\
  &\pm (\nu + \omega_H) \quad  \left( \propto \left( \eta \omega^0_- \right)^2 \right) \
  ,
\label{eq:two_groups_resonances}
\end{align}

as can be seen for the three major peaks in the power spectrum in
Fig.~\ref{fig:two_groups_Scum}. The other peaks are presumably given
by higher resonances. Note that for a circular limit cycle
$\omega^0_+$ or $\omega^0_-$ equals zero leading to vanishing
contributions at $\pm\left( \nu - \omega_H \right)$ or $\pm\left( \nu
  + \omega_H \right)$, respectively.

\begin{figure}[t]
  \centering
  \includegraphics[width=8.5cm]{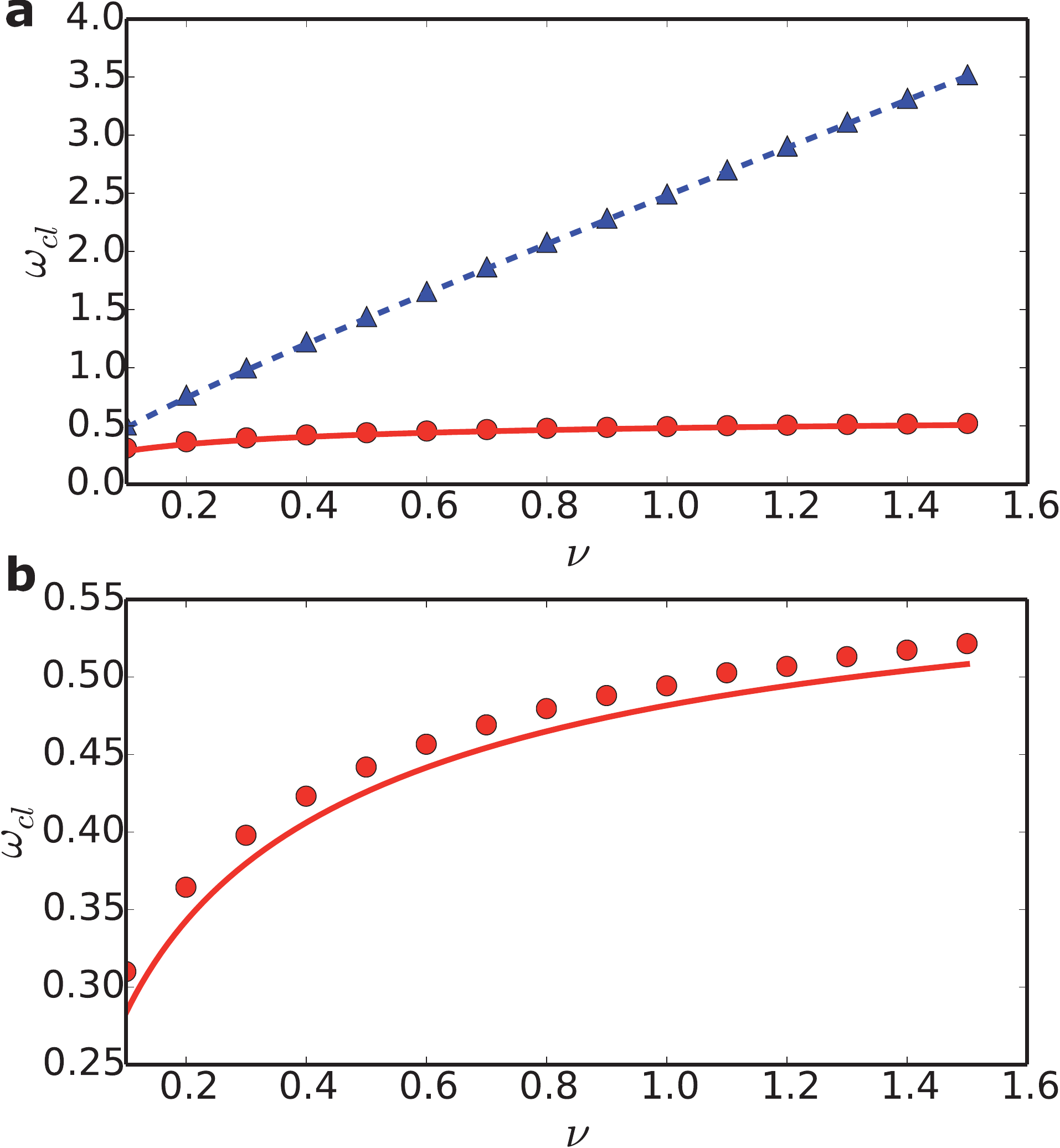}
  \caption{Comparison of the calculated peak frequencies with
  the frequencies in the full system for $c_2 = -0.6$ and $\eta = 0.7$. The Hopf
  bifurcation occurs at $\eta_H = 1/\sqrt 2$. (a) Both frequencies
  $\left| \nu + \omega_H \right|$ in blue (dashed) and $\left| \nu - \omega_H
  \right|$ in red (solid) versus $\nu$. Lines describe the results of the linear
  stability analysis, Eq.~\eqref{eq:two_groups_resonances}, and symbols mark the simulation
  results. (b) For more details only $\left| \nu - \omega_H \right|$
  vs. $\nu$.}
\label{fig:two_groups_cluster_frequencies}
\end{figure}

To further check the validity of the frequencies, obtained via a reduction to
two effective equations and via linear stability analysis, we compare
them with the frequencies in the full system for several values of
$\nu$. The results for $\left| \nu + \omega_H \right|$ (blue, dashed) and
$\left| \nu - \omega_H \right|$ (red, solid) are shown in
Fig.~\ref{fig:two_groups_cluster_frequencies}a. In
Fig.~\ref{fig:two_groups_cluster_frequencies}b we show the comparison
for $\left| \nu - \omega_H \right|$ in more detail.

The simulation results shown are for $\eta = 0.7$, which is close to
the value at the Hopf bifurcation $\eta_H = 1/\sqrt 2 \approx
0.707$. As visible in the figure, the results of the linear stability
analysis (lines), Eq.~\eqref{eq:two_groups_resonances}, reproduce the
simulation results (symbols) very well. The nearly constant shift visible in
Fig.~\ref{fig:two_groups_cluster_frequencies}b is due to the finite
distance to the Hopf bifurcation.

We conclude that the modulated amplitude clusters arise through a Hopf bifurcation
in the rotating frame with frequency $\nu$, which gives rise to the
amplitude modulations in the full system. The dynamics on the created
limit cycle are in anti-phase as to fulfill the conservation
law, which is also in line with our symmetry considerations above. 
Since the Hopf bifurcation occurs in the rotating frame, it is in fact a
secondary Hopf bifurcation. The dynamics in the original frame is thus
quasiperiodic. This is also obvious from the continuous frequency
curves in Fig.~\ref{fig:two_groups_cluster_frequencies}.

\section{Amplitude clusters in the two-groups reduction}

The modulated amplitude clusters described in the preceding section
arise for certain parameters through a Hopf bifurcation. This motion
on a torus can be destroyed through a saddle-node bifurcation leading
to the amplitude clusters shown in Fig.~\ref{fig:clusters}b. These
amplitude clusters are solutions of
Eq.~\eqref{eq:two_groups_reduction_w} in the form $w_1 = R
\exp\left( i\chi_\pm \right)$
\cite{CoulletEmilsson_PhysicaD_1992}, as this results in $\left| W_1
\right| = \eta \sqrt{1 + 2R \cos \chi_\pm + R^2}$. With $\chi_+ =
\chi_- + \pi$ the two solutions describe limit cycles with different radii. Inserting this ansatz into
Eq.~\eqref{eq:two_groups_reduction_w}, separating real and imaginary
parts and assuming $R \neq 0$ one obtains

\begin{align}
  \mu - \eta^2 R^2 - \eta^2 \cos 2\chi - c_2 \eta^2 \sin 2\chi &=0
  \ , \notag \\
  \beta - c_2 \eta^2 R^2 - c_2 \eta^2 \cos 2\chi + \eta^2 \sin 2\chi
  &=0 \ .
\end{align}

This set of equations can be solved for $R$ and $\chi$ and one finds
two pairs of solutions \cite{YochelisSwinney_SIADS_2002, Yochelis_PhDTh}:

\begin{align}
  R^{(1)} &= \sqrt{\frac{\mu + c_2 \beta - \sqrt{\eta^4\left( 1+c_2^2
        \right)^2 - (c_2 - \nu)^2}}{\eta^2 \left( 1 + c_2^2 \right)}}
  \ , \notag \\
  \chi^{(1)}_- &= \frac 12 \arcsin \left( \frac{c_2 - \nu}{\eta^2 \left(
        1+c_2^2 \right)} \right) \ , \notag \\
  \chi^{(1)}_+ &= \chi^{(1)}_- + \pi \ , \\ \notag \\
  R^{(2)} &= \sqrt{\frac{\mu + c_2 \beta + \sqrt{\eta^4\left( 1+c_2^2
        \right)^2 - (c_2 - \nu)^2}}{\eta^2 \left( 1 + c_2^2 \right)}}
  \ , \notag \\
  \chi^{(2)}_- &= \frac \pi 2 - \frac 12 \arcsin \left( \frac{c_2 - \nu}{\eta^2 \left(
        1+c_2^2 \right)} \right) \ , \notag \\ 
\chi^{(2)}_+ &= \chi^{(2)}_- + \pi \ .
\end{align}

We calculate the boundaries $\eta(c_2,\nu)$ of their existence and
obtain:

\begin{align}
  &R^{(1)}, \ \chi^{(1)}_{\pm} \text{ exists for } \eta > \eta_{SN}
  \land \eta < \eta_c \land \eta < \eta_P^- \ , \\
  &R^{(2)}, \ \chi^{(2)}_{\pm} \text{ exists for }
  \begin{cases}
    \eta > \eta_{SN} \ , & \text{for } \eta < \eta_c \ , \\
    \eta_P^- < \eta < \eta_P^+ \ , & \text{for } \eta > \eta_c \ .
  \end{cases}
\end{align}

$\eta_{SN}(c_2, \nu)$, $\eta_c(c_2,\nu)$ and $\eta_P^\pm (c_2,\nu)$ are given by

\begin{align}
  \eta_{SN} &= \sqrt{\frac{\left| c_2 - \nu \right|}{1+c_2^2}} \ ,
  \notag \\
  \eta_c &= \sqrt{\frac{1+c_2 \nu}{2\left(1+c_2^2 \right)}} \ , \notag
  \\
  \eta_P^\pm &=\sqrt{\frac{2(1+c_2 \nu) \pm \sqrt{4(1+c_2 \nu)^2 -
        3\left( 1+c_2^2 \right) \left( 1 + \nu^2 \right)}}{3 \left(
        1+c_2^2 \right)}} \ .
\label{eq:eta_boundaries}
\end{align}

\begin{figure}[t]
  \centering
  \includegraphics[width=8.5cm]{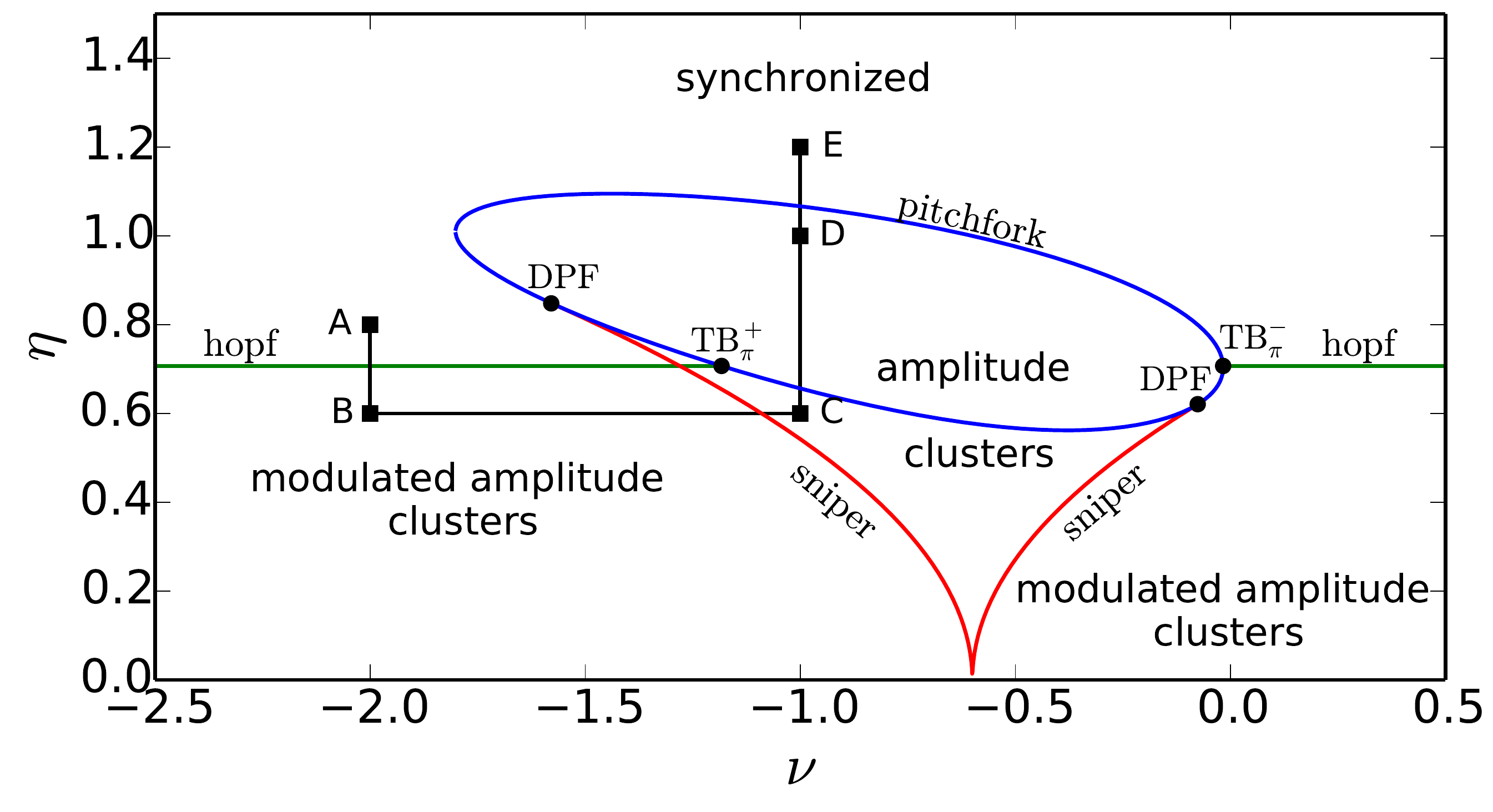}
  \caption{Coarse bifurcation diagram for the two-groups reduction. Shown is
    the parameter space $\eta$ vs. $\nu$ for fixed $c_2 = -0.6$. The
    stable dynamical states are indicated in the figure. The Hopf
    bifurcation (green) is given by $\eta = \eta_H$, the pitchfork
    (blue) is described by $\eta_P^\pm$ and the sniper (red) occurs at $\eta_{SN}$,
    see Eqs.~\eqref{eq:eta_boundaries}. The dynamical states along the
    path A to E are depicted in Fig.~\ref{fig:path_bif_diag}. The
    codimension-two points are two Takens-Bogdanov points of
    $\pi$-rotational symmetry ($\mathrm{TB_\pi^\pm}$) and two
    degenerate pitchfork bifurcations (DPF). The details of the
    bifurcation structure, which have been omitted here, including the
    unfoldings of the $\mathrm{TB_\pi^\pm}$ points, will be discussed
    in Section~\ref{sec:bif_diag_details}.}
\label{fig:eta_boundaries}
\end{figure}

\begin{figure*}[t]
  \centering
  \includegraphics[width=17cm]{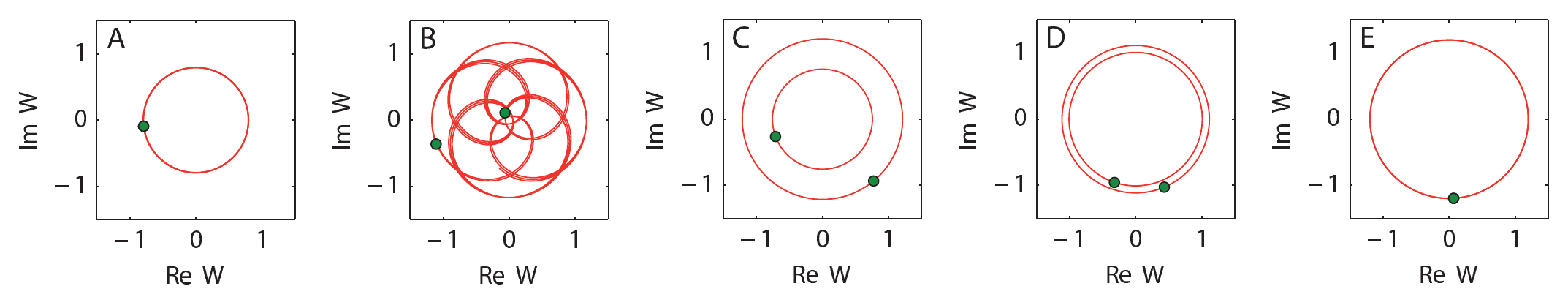}
  \caption{Simulation results for the two-groups reduction in the 
	original frame illustrating the dynamical states along the path A-E 
	in the bifurcation diagram in Fig.~\ref{fig:eta_boundaries}.}
\label{fig:path_bif_diag}
\end{figure*}

Linear stability analysis reveals that the amplitude cluster solutions
$R^{(1,2)}\exp\left(i\chi^{(1,2)}_\pm\right)$ arise as two saddle-node pairs at $\eta_{SN}$,
thereby destroying the limit cycle of the modulated amplitude clusters
in a saddle-node of infinite period bifurcation (sniper). Solution (1)
is a saddle and solution (2) is a stable node. Both solutions (1) and
(2) can be destroyed in a pitchfork bifurcation with the synchronized
solution. For details see the next section. In
essence, the amplitude clusters emerge in a sniper bifurcation when
coming from a parameter region, where the modulated amplitude clusters
are stable. And they arise in a pitchfork
bifurcation when coming from a parameter region, where the
synchronized solution is stable (in a small region they also arise via
a saddle-node bifurcation; see next section). A coarse bifurcation
diagram is depicted in Fig.~\ref{fig:eta_boundaries} with 
illustrations of the dynamical states along the path A to E given in 
Fig.~\ref{fig:path_bif_diag}. 

The overall structure reminds of a 
so-called Arnold tongue and we will discuss the relation to the 
locking behaviour of forced oscillatory media in 
Section~\ref{sec:conclusions}. Inside the tongue one observes 
amplitude clusters. The tongue is bounded by a sniper bifurcation 
for small $\eta$ values and by a pitchfork bifurcation for high 
$\eta$ values. A hopf bifurcation separates the region of modulated 
amplitude clusters from the region of stable synchronized solutions. To 
illustrate the different dynamical behaviors in the distinct regions, we go through the path A to E (for comparison see 
Fig.~\ref{fig:path_bif_diag}): Starting at A with 
the synchronized solution, the hopf bifurcation creates the limit 
cycle for the modulated amplitude clusters in B. This limit cycle is then 
destroyed by the sniper bifurcation resulting in amplitude clusters 
in C. Approaching the outer pitchfork bifurcation brings the 
fixed points of the amplitude clusters closer together in D. 
At the pitchfork the fixed points of the amplitude clusters merge 
with the synchronized solution with what we end up in E. Note that, 
as $w_2 = -w_1$ in Eq.~\eqref{eq:two_groups_reduction_w}, both 
groups undergo the bifurcations simultaneously and the second group 
always realizes the $\pi$-rotated solution of the first group.

Furthermore we encounter three codimension-two bifurcations, namely 
a degenerate pitchfork (DPF) and two types of Takens-Bogdanov points 
$\mathrm{TB_\pi^\pm}$. The unfoldings
of the Takens-Bogdanov points are presented in the next Section. Note
that due to the symmetry present in the system, the unfoldings are much
more complicated than in the standard case.

This diagram is strictly valid only for the two-groups reduction. It
clarifies, which bifurcations lead to the amplitude and modulated
amplitude clusters. The diagram is applicable whenever the full
ensemble is separated into two subgroups.

\section{Details of the bifurcation diagram}
\label{sec:bif_diag_details}



\begin{figure}[ht]
  \centering
  \includegraphics[width=8.5cm]{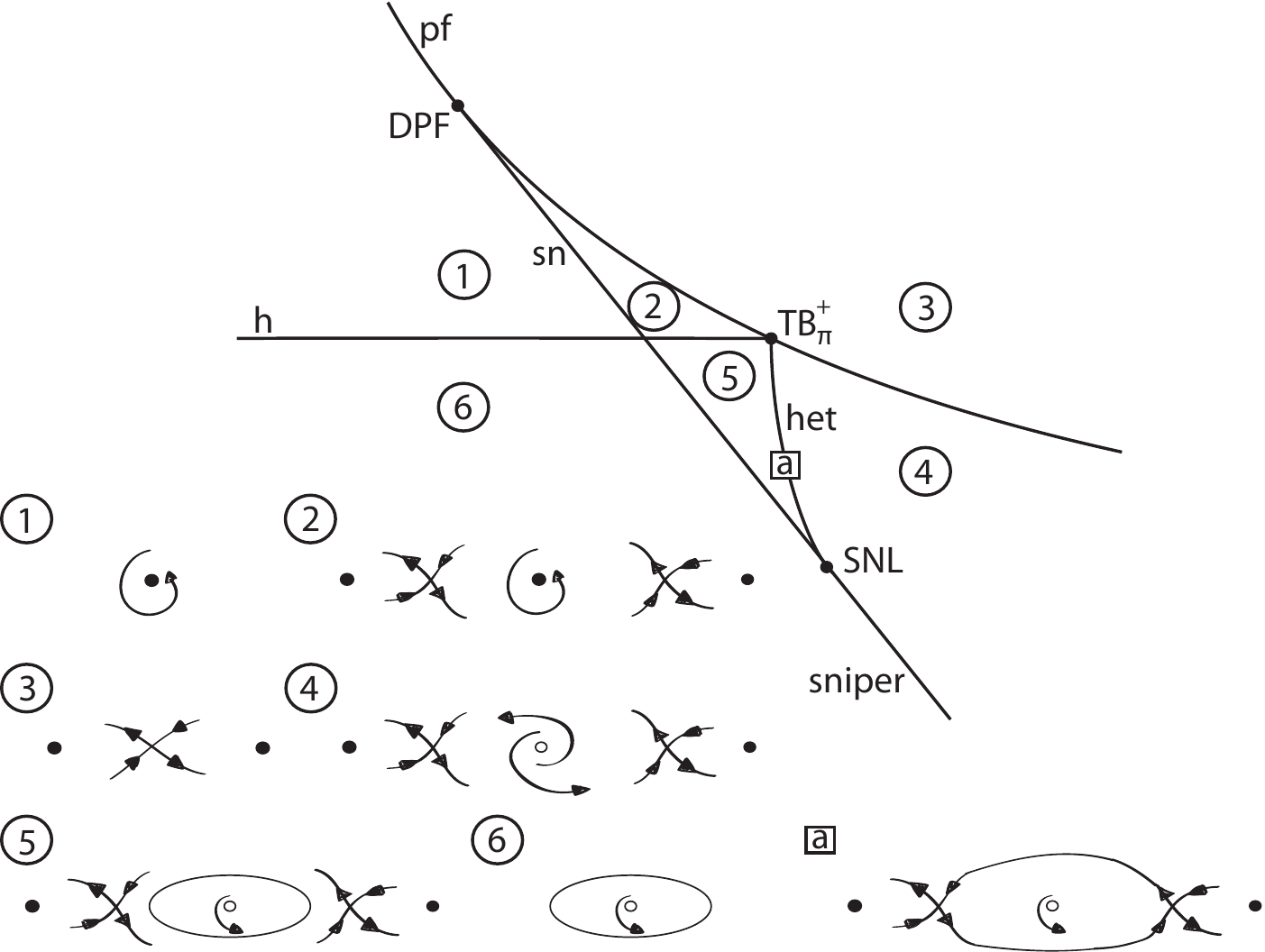}
  \caption{Sketch of the local bifurcation structure around the $\mathrm{TB_\pi^+}$ point
    with corresponding phase portraits. The involved codimension-one bifurcations are:
    pitchfork (pf), saddle-node (sn), Hopf (h), saddle-node of infinite
    period (sniper) and heteroclinic (het). The codimension-two bifurcations
    are: Takens-Bogdanov $\mathrm{TB_\pi^+}$, degenerate pitchfork (DPF)
    and saddle-node loop (SNL). Stable fixed points are marked by
    filled circles and unstable ones by empty circles. Stable limit
    cycles are drawn with a solid line and unstable limit cycles with
    a dashed line.}
  \label{fig:tb_plus}
\end{figure}
\begin{figure}[ht]
  \centering
  \includegraphics[width=8.5cm]{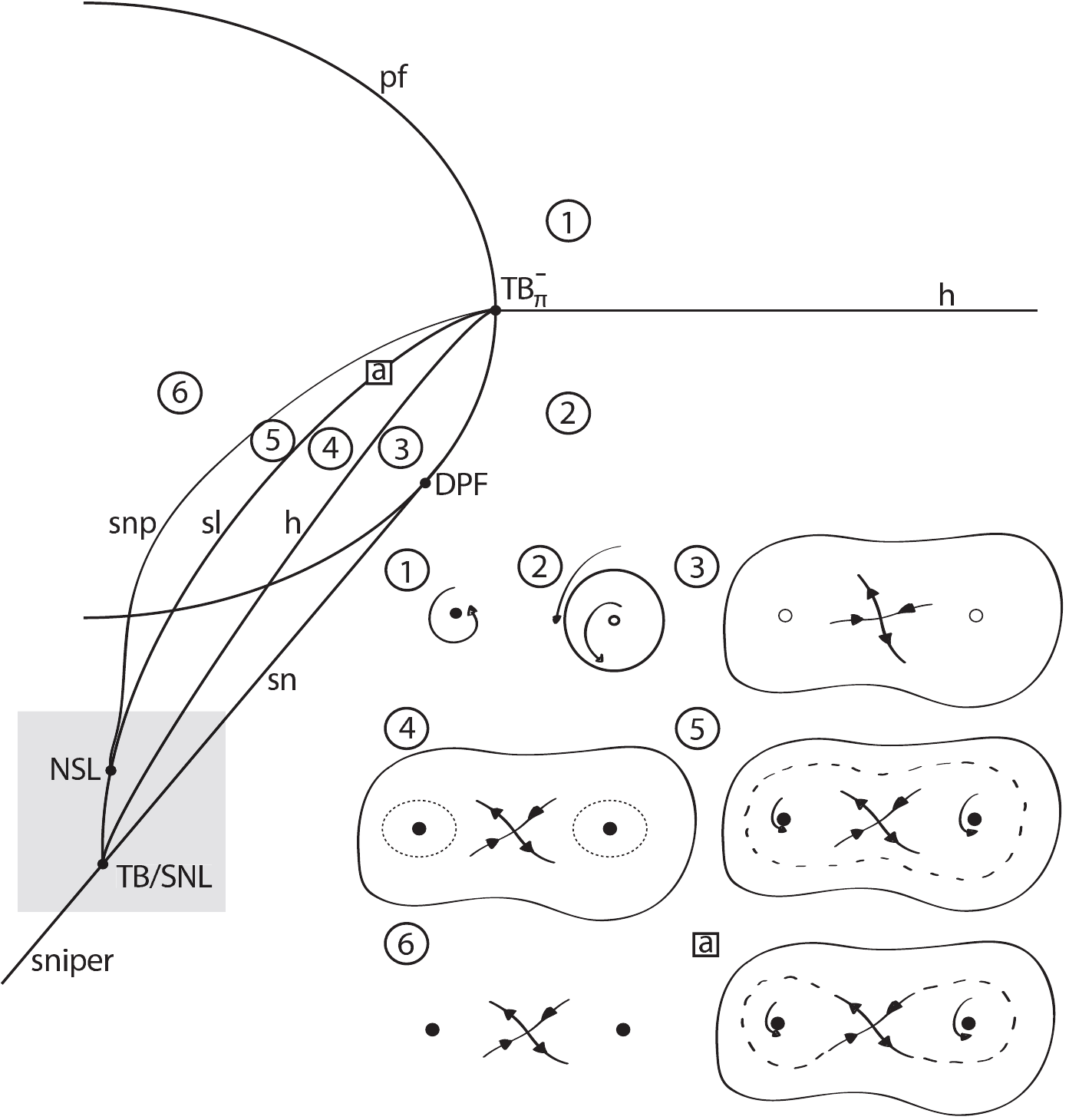}
  \caption{Sketch of the local bifurcation structure around the $\mathrm{TB_\pi^-}$ point
    with corresponding phase portraits. The involved codimension-one bifurcations are:
    pitchfork (pf), saddle-node (sn), Hopf (h), saddle-node of infinite
    period (sniper), saddle-loop (sl) and saddle-node of periodic orbits
    (snp). The codimension-two bifurcations
    are: Takens-Bogdanov with symmetry ($\mathrm{TB_\pi^-}$) and without
    symmetry (TB), saddle-node loop (SNL), degenerate pitchfork (DPF)
    and neutral saddle-loop (NSL). Here, a TB and a SNL belonging to
    different solutions coincide, for details see text. Stable fixed points are marked by
    filled circles and unstable ones by empty circles. Stable limit
    cycles are drawn with a solid line and unstable limit cycles with
    a dashed line. Note that the bifurcation structure in the shaded
    box is not a result of the continuation as this diverges. It is
    consistent with the rest of the diagram, but there might be other
    bifurcations involved, see e.g. Ref.~\cite{Vance_JCP_1989}.}
  \label{fig:tb_minus}
\end{figure}

The codimension-two bifurcations $\mathrm{TB^\pm_\pi}$ present 
in the coarse bifurcation diagram in Fig.~\ref{fig:eta_boundaries} 
have rather complex unfoldings. Using the software 
\texttt{AUTO-07P} for numerical continuation, we could identify the 
local and global bifurcations occurring around the 
$\mathrm{TB_\pi^\pm}$ points. 
The unfolding of the plus case, $\mathrm{TB_\pi^+}$, is shown 
schematically in
Fig.~\ref{fig:tb_plus}, while the minus case, $\mathrm{TB_\pi^-}$, is
presented in Fig.~\ref{fig:tb_minus}. Sketches of corresponding phase portraits 
are also depicted in the figures.

In the $\mathrm{TB_\pi^+}$ point a pitchfork, a Hopf and a
heteroclinic bifurcation meet. In our system, we find in the vicinity
also a saddle-node bifurcation, which meets the pitchfork in a
degenerate pitchfork bifurcation (DPF) and the heteroclinic in a
saddle-node loop (SNL) bifurcation, see Fig.~\ref{fig:tb_plus}. The
DPF turns the pitchfork from supercritical to subcritical and the
$\mathrm{TB_\pi^+}$ changes it back to supercritical. The SNL turns
the saddle-node into a saddle-node of infinite period (sniper). 

When following the numbering in Fig.~\ref{fig:tb_plus}, we start with a
stable focus (1), then cross the saddle-node, thereby creating two saddle
node pairs (2). Then, we cross the subcritical pitchfork and end up in
(3) with two stable nodes and a saddle. Next, we cross the pitchfork
on the supercritical side, yielding two saddle node pairs with an
unstable focus in between (4). Note that the foci involved in 
the pitchfork bifurcations change to nodes just before the 
bifurcations occur. Crossing the heteroclinic bifurcation creates a
stable limit cycle around the unstable focus in the center (5), which
emerges from a double heteroclinic connection at the bifurcation
(a). Finally, the saddle node pairs are annihilated in a saddle-node
bifurcation and we are left with a stable limit cycle around an
unstable focus in (6).

The local bifurcation structure around the $\mathrm{TB_\pi^-}$ point
is more complex. In the $\mathrm{TB_\pi^-}$ point, a pitchfork, a Hopf
concerning the synchronized solution, a Hopf concerning the amplitude
cluster solutions, a saddle-loop and a saddle-node of periodic orbits
(snp) meet. The saddle-loop line is in fact the coincidence of two
saddle-loop bifurcations, one which describes the saddle-loop
bifurcation of the amplitude cluster solutions (small limit cycles in
Fig.~\ref{fig:tb_minus}) and one which concerns the modulated
amplitude cluster solutions (outer limit cycles in
Fig.~\ref{fig:tb_minus}). With this, we can understand the
codimension-two bifurcations occurring in the vicinity of the
$\mathrm{TB_\pi^-}$ point: the snp and the two saddle-loops meet first in a
neutral saddle-loop (NSL) and later the saddle-loops meet with the hopf
and the saddle-node in a Takens-Bogdanov (TB) without symmetry and a
saddle-node loop (SNL). The saddle-loop corresponding to the amplitude
cluster solution ends in the TB point and the other saddle-loop turns
the saddle-node into a saddle-node of infinite period (sniper) at the
SNL. Note that this region of the bifurcation diagram, i.e. the shaded
region, is not a result of the continuation as this diverges. It is
consistent with the rest of the diagram, but there might be other
bifurcations involved, see e.g. Ref.~\cite{Vance_JCP_1989}.
In the degenerate pitchfork (DPF) the saddle-node bifurcation
meets the pitchfork.

Again we can go through the diagram step by step by following the
numbering in Fig.~\ref{fig:tb_minus}: We start with a stable focus (1)
and cross the Hopf to obtain a stable limit cycle around an unstable
focus (2). Then, the subcritical pitchfork turns the unstable focus
into a saddle point and creates two unstable nodes (3). The
subcritical hopf creates two unstable limit cycles (4), which form
homoclinic loops when meeting the manifolds of the saddle point in the
saddle-loop bifurcation (a). This saddle-loop bifurcation coincides
with a saddle-loop bifurcation of an unstable modulated amplitude
cluster solution, which is given by the unstable limit cycle in
(5). Finally, the stable and the unstable limit cycle annihilate each
other in a snp, and a pair of stable nodes (describing the
amplitude cluster solutions) with a saddle point in
between remain (6).

In fact the $\mathrm{TB^\pm_\pi}$ points are Takens-Bogdanov points 
of $\pi$-rotational or cubic symmetry \cite{GuckenheimerHolmes_1983, 
Guckenheimer_JMA_1984}. This is the symmetry present in 
Eq.~\eqref{eq:two_groups_reduction_w}. They possess the same 
principal bifurcation structure as the second order resonance points found in 
the investigation of periodically forced oscillators 
\cite{Vance_JCP_1989}. However, some bifurcations are different, as 
we will discuss in the next section.

\section{Conclusions}
\label{sec:conclusions}

We could unravel the complex bifurcation structure exhibited by the 
two-cluster solutions of an ensemble of generic limit-cycle 
oscillators near a Hopf bifurcation. The conservation of the 
mean-field oscillations leads to mainly two bifurcations: a Hopf bifurcation yielding the 
modulated amplitude clusters and a pitchfork bifurcation resulting 
in common amplitude clusters. The meeting of these two gives rise to 
two Takens-Bogdanov points of $\pi$-rotational symmetry and 
therewith to a wide variety of dynamical states.

Besides the application to the experimental system, for which the 
model was originally proposed, namely the photoelectrodissolution of 
n-type silicon \cite{Miethe_PRL_2009, GarciaMorales_PRE_2010, 
Schoenleber_NJP_2014}, there is a strong connection to resonantly 
forced oscillatory media \cite{Gambaudo_JDE_1985, 
CoulletEmilsson_PhysicaD_1992, Petrov_Nature_1997, 
GarciaMorales_ContempPhys_2012, Lin_PRE_2004, Yochelis_EPL_2005, 
Kaira_PRE_2008, Conway_PRE_2007, Marts_Chaos_2006, 
Yochelis_PhysicaD_2004}. The symmetry properties of the reduced 
dynamics in Eq.~\eqref{eq:two_groups_reduction_w}, namely 
the cubic and $\pi$-rotational symmetries, are also present in
the complex Ginzburg-Landau equation (CGLE) with resonant forcing near a 2:1 resonance. In fact, there is a linear
transformation that transforms the equation for $w_1$ in
Eq.~\eqref{eq:two_groups_reduction_w} to the form given in
e.g. Ref.~\cite{CoulletEmilsson_PhysicaD_1992} (see Eq. (10) therein) of
the resonantly forced CGLE, when omitting the diffusive coupling. As for forced
oscillatory media, we observe an Arnold tongue, a region of
frequency locking, in the bifurcation diagram in
Fig.~\ref{fig:eta_boundaries}. The tongue starts at $\nu = c_2$, 
i.e. at a value of the driving frequency $\nu$ equal to the 
natural frequency of the Stuart-Landau oscillator $c_2$. The locking region is bounded by
the saddle-node, sniper and pitchfork bifurcations. The dynamics lock
to the frequency $\nu$ of the mean-field oscillations, i.e. to the 
frequency of the driving. Thus, we observe an 1:1 locking instead of 
a 2:1 locking, which one would expect, since we observe the 
bifurcation structure of a 2:1 resonance.
This is reflected in the occurrence of a pitchfork 
bifurcation instead of the period doubling bifurcation, see 
Ref.~\cite{Vance_JCP_1989}. Furthermore, as in the forced CGLE, the locked solutions do not
lie on a torus, since the torus is destroyed in a sniper bifurcation.

In our system the forcing
is in fact a self-forcing, as the dynamics produce a
mean-field oscillation, which is conserved and then acts back as a
forcing on the system. This self-forcing renders the cluster 
solutions possible. But note that it is the mathematical structure 
of a 2:1 resonance that is responsible for the cluster formation. We observe
an 1:1 locking and in general this would not give rise to cluster 
formation.

\begin{acknowledgments}
We thank Vladimir Garc\'{i}a-Morales for fruitful discussions, his
comments on the symmetry considerations and for careful reading of the manuscript.
The authors gratefully acknowledge financial support from the
\textit{Deutsche Forschungsgemeinschaft} (Grant
no. KR1189/12-1), the \textit{Institute for Advanced Study, Technische
  Universit\"{a}t M\"{u}nchen} funded by the German Excellence
Initiative and the cluster of excellence \textit{Nanosystems
  Initiative Munich (NIM)}.
\end{acknowledgments}

\bibliography{lit}

\end{document}